\documentclass[epj]{svjour}
\usepackage{graphics}

\newcommand{\ba}{\begin{array}}
\newcommand{\ea}{\end{array}}
\newcommand{\be}{\begin{equation}}
\newcommand{\ee}{\end{equation}}
\newcommand{\bea}{\begin{eqnarray}}
\newcommand{\eea}{\end{eqnarray}}
\newcommand{\beq}{\begin{equation}}
\newcommand{\eeq}{\end{equation}}
\def\npb#1#2#3{    { Nucl. Phys. }{\bf B #1} (#2) #3}

\def\prd#1#2#3{    { Phys. Rev. }{\bf D #1} (#2) #3}

\begin{document}
\title{New NNLL QCD Results on the Decay $B \rightarrow X_s \ell^+ \ell^-$\thanks{Contribution to International Europhysics Conference on High Energy Physics EPS03, 17-23 July, Aachen, Germany, presented by T.H.}}
\author{ A.~Ghinculov\inst{1} \and T.~Hurth\thanks{{Heisenberg Fellow}}\inst{2}
\and  G.~Isidori\inst{3}  \and Y.-P.~Yao \inst{4}}

\institute{Department of Physics, University of Rochester, Rochester, NY 14627, USA \and Theoretical Physics Division, CERN, CH-1211 Gen\`eve 23, Switzerland
 and\\ SLAC, Stanford University, Stanford, CA 94309, USA \and INFN, Laboratori Nazionali di Frascati, I-00044 Frascati, Italy \and 
Michigan Center for Theoretical Physics, Univ. of Michigan, Ann Arbor 
 MI 48109-1120, USA}

\date{{\tt CERN-TH/2003-249}, {\tt SLAC-PUB-10208}, {\tt PITHA 03/09}, {\tt MCTP-03-46}}

\abstract{We present here new NNLL predictions on the inclusive rare
decay $B \rightarrow X_s \ell^+ \ell^-$ based on our new two-loop QCD analysis
of the four-quark operators.}


\authorrunning{A.Ghinculov,T.Hurth,G.Isidori,Y.P.Yao}
\titlerunning{New NNLL Results on the Decay $B \rightarrow X_s \ell^+ \ell^-$}

\maketitle

\section{Introduction}
\label{intro}

The rare decay $B \rightarrow X_s \ell^+ \ell^-$ offers the
$B$ factories the possibility to open a new interesting 
window on flavour physics.  
A measurement of the dilepton spectrum and of the forward--backward
asymmetry in this decay process has been proposed as a way to test for new physics and 
 discriminating  between different new physics scenarios (for a recent review see \cite{hurth}).
Indeed, the short-distance-dominated  flavour-changing neutral-current
amplitude of $B \rightarrow X_s \ell^+ \ell^-$ is extremely sensitive to 
possible new degrees of freedom, even if these appear well above the electroweak scale. 
This type of indirect search for new physics relies both on an accurate measurement of the 
dilepton spectrum, and on an accurate 
theoretical calculation of the decay probability.
The inclusive $B \rightarrow X_s \ell^+\ell^-$ transition
just starts to be accessible at the $B$ factories:
BELLE and BABAR have already 
data on the rate based on  a semi-inclusive analysis 
\cite{BELLEbsll2,BELLEbsll3,Babarbsll}.  
These first measurements are in agreement with SM expectations,
but are still affected by a $30 \%$ error: substantial improvements 
can be expected in the near future.

\begin{figure}[t]
\begin{center}
\resizebox{0.5 \textwidth}{!}{%
  \includegraphics{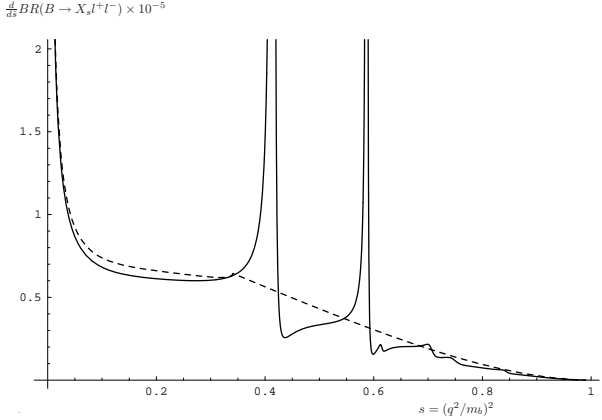} }
\vspace{-1.0 cm}
\end{center}
\caption{Schematic dilepton mass spectrum of $B \rightarrow X_s \ell^+ \ell^-$,
the dashed line corresponds to the perturbative contribution}
\label{fig:1}   
\end{figure}

From the theoretical point of view, inclusive rare decay modes like 
$B \rightarrow X_s \ell^+ \ell^-$ are very attractive because, 
in contrast to most of the exclusive channels, they 
are theoretically clean observables dominated by the 
partonic contributions. Non-perturbative effects 
in these transitions are small and can be 
systematically accounted for, through 
an expansion in inverse powers of the heavy $b$ quark mass.
In the specific case of  $B \rightarrow X_s \ell^+ \ell^-$, 
the latter statement is applicable only if the
$c \bar c$ resonances that show up as large peaks in the dilepton 
invariant mass spectrum (see fig. \ref{fig:1}) are removed by
appropriate kinematic cuts. In the {\em perturbative windows}, namely  
in the region below and in the one above the resonances,
theoretical predictions for the invariant mass spectrum
are dominated by the purely perturbative contributions, 
and a theoretical precision comparable with  the one reached  
in the decay $B \rightarrow X_s \gamma$ is in principle possible
\cite{nonpert,nonpert2}.
Regarding the choice of precise cuts in the dilepton mass 
spectrum, it is important to stress that the maximal precision 
is obtained when theory and experiments are compared employing 
the same energy cuts and avoiding any kind of extrapolation.

In these processes QCD corrections lead to a sizable modification of the 
pure short-distance electroweak contribution, generating large logarithms 
of the form $\alpha_s^n(m_b)$ $\times\log^m(m_b/M_{\rm heavy})$,
where $M_{\rm heavy}=O(M_W)$ and $m \le n$ (with $n=0,1,2,...$).
These effects  are induced by 
hard--gluon exchange between the quark lines of the one-loop electroweak
diagrams.
The most suitable framework for their necessary resummations 
is an effective low-energy 
theory with five quarks, obtained by integrating out the
heavy degrees of freedom. 
Renormalization-group (RG) 
techniques allow  for the resummation of  the series of 
leading logarithms (LL),  $\alpha_s^n(m_b) \,  \log^n(m_b/M)$,
next-to-leading logarithms (NLL), $\alpha_s^{n+1}(m_b) \, \log^n (m_b/M)$, and so on.

\section{NNLL Calculation}
\label{nnll}

For a detailed discussion of the present status of the perturbative 
contributions to decay rate and 
FB asymmetry of $B \to X_s \ell^+\ell^-$ 
we refer to \cite{hurth}. Here we simply recall that 
the complete NLL contributions to the decay amplitude
can be found in \cite{MM}. 
Since the LL contribution to the rate turns out to be numerically 
rather small, NLL terms represent an $O(1)$ correction and 
a computation of NNLL terms is needed 
if we aim for a numerical accuracy below $10 \%$, similar to 
the one achieved by the NLL calculation of $B \to X_s \gamma$.
Large parts of the latter can be taken over and used in the NNLL 
calculation of $B \rightarrow X_s \ell^+ \ell^-$. Thanks to the 
joint effort of several groups \cite{MISIAKBOBETH,Gambino,Asa1,Asa2,Adrian1,Asa3,Adrian2,newnew}, 
the  necessary {\it additional} NNLL calculations have been practically 
finalized  by now.
 
The computation of all the missing initial conditions to NNLL 
precision has been presented in Ref.~\cite{MISIAKBOBETH}. 
The most relevant missing piece of the anomalous dimension matrix, 
namely the three-loop mixing of four-quark operators  
into semileptonic ones, has been obtained very recently in Ref.~\cite{Gambino}. 
Thanks to these works, the large matching scale uncertainty 
of around $16 \%$ of the NLL prediction has been removed.

The two-loop matrix elements of the four-quark operators 
are probably the most difficult part of the NNLL enterprise. 
Asatrian et al. succeeded in using  
a mass and momentum double expansion of the virtual 
two-loop diagrams \cite{Asa1}. 
This calculation is based on expansion techniques
which can be applied only in the lower perturbative window of the 
 dilepton spectrum.
Once the $c \bar c$ threshold is reached, the momentum expansion
is not a valid procedure  anymore.
We have recently extended this calculation to the upper perturbative 
window above the $c \bar c$ threshold \cite{Adrian2,newnew}: in the next section  
we shall present the first phenomenological outcomes of this analysis.
We resorted to semi-numerical methods,  which are valid both below and above the threshold. 
Regarding the lower perturbative window, our analysis provides as an independent confirmation
of \cite{Asa1}, which is particularly welcome in view of its complexity:
below the  $c \bar c$ threshold the agreement between our results
and those of \cite{Asa1}  is  excellent. 
As shown in \cite{Asa1}, the NNLL  matrix element contributions are a
fundamental ingredient to reduce the perturbative uncertainty:
in the lower window the (low) scale dependence gets reduced  
from $13\%$ to $6.5\%$. As we shall discuss in the next
section, the residual scale dependence in the upper window is even 
smaller, around the $3\%$ level.

Another independent ingredient of the NNLL analysis is represented 
by bremsstrahlung corrections (and corresponding virtual soft-gluon terms).
Also this part of the calculation is now available and cross-checked 
for both dilepton spectrum and FB asymmetry \cite{Asa1,Asa2,Adrian1,Asa3}.
While the NNLL program for the FB asymmetry is fully completed \cite{Adrian1,Asa3},
in principle there are some pieces still missing for the 
integrated dilepton spectrum; however, all of them are estimated to 
be below $1 \%$  at the branching ratio level. Note that, at this 
level of accuracy, other subleading effects become more important.
For instance, further studies regarding higher-order 
electromagnetic effects should deem  necessary.

\section{Results}
\label{sec:results}

Before discussing the numerical predictions for the integrated 
branching ratios, it is worth to emphasize that low- and high-dilepton mass regions 
have complementary vir\-tues and disadvantages.
These  can be summarized as follows ($q^2=M_{\ell^+\ell^-}^2$): 
\begin{enumerate}
\item[] 
{\em Virtues of the low-$q^2$ region:} reliable $q^2$ spectrum; 
small $1/m_b$ corrections; sensitivity to the interference of $C_7$
and $C_9$; high rate. 
\item[]
{\em Disadvantages of the low-$q^2$ region:}  difficult to perform 
a fully inclusive measurement (severe cuts on the dilepton 
energy and/or the hadronic invariant mass); long-distance effects 
due to processes of the type $B \to \Psi X_s \to  X_s + X^\prime \ell^+\ell^-$ 
not fully under control; non-negligible scale 
and $m_c$ dependence. \\
\item[]
{\em Virtues of the high-$q^2$ region:} negligible scale 
and $m_c$ dependence due to the strong sensitivity to the Wilson 
coefficient $|C_{10}|^2$; 
easier to perform a fully inclusive measurement (small 
hadronic invariant mass); negligible long-distance effects of the type 
$B \to \Psi X_s \to  X_s + X^\prime \ell^+\ell^-$. 
\item[]
{\em Disadvantages of the high-$q^2$ region:} 
$q^2$ spectrum  not reliable; sizable $1/m_b$ corrections; low rate.
\end{enumerate}

\noindent
Given this situation, we believe that future experiments should 
try to measure the branching ratios in both regions and 
report separately the two results. These two measurements 
are indeed affected by different systematic uncertainties 
(of theoretical nature) but they  provide different 
short-distance information. 

In order to obtain theoretical predictions that  can be confronted with
experiments, it is necessary to convert the $s=q^2/m_b^2$ range into 
a range for the measurable dilepton invariant mass $q^2$. Concerning the 
 low-$q^2$ region, we propose as reference interval the range
$q^2 \in [1,6]~{\rm GeV}^2$.
The lower bound on $q^2$ is imposed in order to cut a region where 
there is no new information with respect to $B\to X_s \gamma$ 
and where we cannot trivially combine electron and muon modes.
Taking into account the input values in Table~\ref{tab:main_inputs}, 
the NNLL prediction within the SM is:

\renewcommand\arraystretch{1.3}
\begin{table}[t]
\centering
\begin{tabular}{|c|c|}
\hline
   $m^{\rm pole}_b=(4.9\pm 0.1)$~GeV    
&  $m_c/m_b=0.29 \pm 0.02$
 \\ \hline
   $\alpha_s(M_Z)=0.119$ 
&  $\alpha_{\rm em}=1/128$ 
 \\ \hline
 $\mu=\left(5.0^{+5.0}_{-2.5}\right)$~GeV  
& $|V_{ts}/V_{cb}|=0.97$
 \\ \hline
\end{tabular}
\caption{Main input values used in the numerical analysis.}
\label{tab:main_inputs}
\end{table}
\renewcommand\arraystretch{1}

\begin{eqnarray}
R^{\rm low}_{\rm cut} &=&
 \int_{ 1~{\rm GeV}^2 }^{6~{\rm GeV}^2 } dq^2 
\frac{ d \Gamma(B\to X_s \ell^+\ell^-)}{\Gamma(B\to X_c e \nu)}
= 1.48 \times 10^{-5} \nonumber \\
 &\times&   
\Biggl[ 1  \pm 8\% \big|_{\Gamma_{\rm sl}}  \pm 6.5\% \big|_{\mu}   \pm 2\% \big|_{m_c}    
\pm 3\% \big|_{m_b({\rm cuts})} \nonumber \\
&&+ (4.5 \pm 2)\% \big|_{1/m^2_b}  - (1.5 \pm 3)\% \big|_{c\bar c}  \Biggl] \nonumber \\
 &=&   (1.52 \pm 0.18 )\times 10^{-5}~. 
\end{eqnarray}

The error denoted by $\Gamma_{\rm sl}$ corresponds to the theoretical 
uncertainty implied by the $\Gamma(B\to X_c e \nu)$ normalization
which, in turn, is dominated by the uncertainty on $m_c$. As already 
stressed in Ref.~\cite{Asa1}, at present this is the dominant source
of uncertainty in the low-$q^2$ region. 
In principle, alternative normalizations 
such as the one proposed in Ref.~\cite{MG} using 
$B\to X_u \ell \nu$ could be used to 
reduce this uncertainty in the future; however, in practice 
this is still the best we can do at the moment. Using the world average 
$\Gamma(B\to X_c e \nu)=(10.2\pm0.4)\%$, we finally obtain:
\begin{equation}
BR(q^2 \in [1,6]~{\rm GeV}^2 ) = (1.55 \pm 0.19) \times 10^{-6}~.
\end{equation}

Concerning the high-dilepton mass region, we propose as a reference 
cut $q^2 > 14.4 {\rm GeV}^2$. Using our new NNLL evaluation 
of the two-loop matrix elements (and reanalyzing nonperturbative 
effects in this region) we find:
\begin{eqnarray}
R^{\rm high}_{\rm cut} &=&
 \int_{ q^2> 14.4~{\rm GeV}^2 } dq^2 
\frac{ d \Gamma(B\to X_s \ell^+\ell^-)}{\Gamma(B\to X_c e \nu)} = \nonumber \\
&=&  4.09 \times 10^{-6}  \times 
 \Biggl[ 1  \pm 8\% \big|_{\Gamma_{\rm sl}}  \pm 3\% \big|_{\mu}  \nonumber \\
&& \pm 15\% \big|_{m_b({\rm cuts})} 
- (8 \pm 8)\% \big|_{1/m^{(2,3)}_b}  \pm 3 \% \big|_{c\bar c}  \Biggl] \nonumber \\
&=&   (3.76 \pm 0.72 )\times 10^{-6}~,\quad 
\end{eqnarray}
or 
\begin{equation}
BR(q^2 > 14.4~{\rm GeV}^2 ) = (3.84 \pm 0.75) \times 10^{-7}~.
\end{equation}
As can be noted, in this case $\mu$ dependence  
and intrinsic $m_c$ dependence induce a negligible 
uncertainty. In addition to the semileptonic normalization
(which induces a common uncertainty to low- and high-$q^2$
regions), here the largest uncertainties are related to 
the $1/m_b$ expansion. Most of them are parametric,  
 which could be reduced in the future,
with the help of more precise data on 
charged-current semileptonic decays 
$B\to X_{u,c} \ell \nu$. The leading $15\%$ error denoted 
by `$m_b({\rm cuts})$' indicates the uncertainty in the 
relation between the physical  $q^2$ interval and the corresponding 
interval for the variable $s$ of the partonic calculation: this is
nothing but the uncertainty in the relation between $m_b$ and 
the physical hadron mass. 
The $8\%$ error denoted by $1/m^{(2,3)}_b$ indicates 
the combined uncertainty due to $1/m_b^2$ and $1/m_b^3$ corrections:
these non-perturbative effects induce a divergence 
in the  $q^2$ spectrum (for $q^2 \to m_b^2$) which cannot be re-absorbed 
in a shape function distribution \cite{nonpert2,bauer}. This divergence 
does 
not  prevent us from  making reliable predictions for the integrated rate, 
but it slows down the convergence of the series, which turns 
out to be an effective expansion in inverse powers 
of $m_b(1-\sqrt{s_{\rm max}})$, rather than $m_b$.\footnote{For a more 
detailed discussion of these results we refer the reader to \cite{newnew}.}

\medskip 

We finally note that in both cases (low- and high-$q^2$
regions), we have not explicitly indicated the uncertainty 
due to $\alpha_{\rm em}$ and $|V_{ts}/V_{cb}|$,
which have been fixed to the values in Table~\ref{tab:main_inputs}.
In principle, variations of these parameters can be trivially  
taken into account by appropriate multiplicative factors (they both appear 
as a squared multiplicative factors in $R_{\rm cut}$).
However, we stress that a coherent treatment of higher-order 
electromagnetic effects --- which is beyond the scope of 
this work --- cannot be simply reabsorbed into a redefinition of 
$\alpha_{\rm em}$.

\end{document}